\documentclass[aps,pra,twocolumn,amsmath,showpacs,
amssymb,amsfonts]{revtex4}

\usepackage{graphics}
\usepackage{epsfig}
\usepackage{amsmath}
\usepackage{amsthm}

\begin{document}
\title{Quantum Private Queries: security analysis}

\author{Vittorio~Giovannetti$^1$,
        Seth~Lloyd$^2$, 
        and~Lorenzo~Maccone$^3$, }
        
\address{$^1$NEST-CNR-INFM \& Scuola Normale
  Superiore, Piazza dei Cavalieri 7, I-56126, Pisa, Italy e-mail: v.giovannett@sns.it \\
$^2$MIT, RLE and Dept. of Mech. Engin. MIT 3-160, 77 Mass.~Av., Cambridge, MA
  02139, USA \\
$^3$QUIT, Dip. Fisica ``A. Volta'', Univ. Pavia, via
  Bassi 6, I-27100 Pavia, Italy}


\begin{abstract}
  We present a security analysis of the recently introduced Quantum
  Private Query (QPQ) protocol. It is a cheat sensitive quantum
  protocol to perform a private search on a classical database. It
  allows a user to retrieve an item from the database without
  revealing which item was retrieved, and at the same time it ensures
  data privacy of the database (the information that the user can
  retrieve in a query is bounded and does not depend on the size of
  the database). The security analysis is based on
  information-disturbance tradeoffs which show that whenever the
  provider tries to obtain information on the query, the query
  (encoded into a quantum system) is disturbed so that the person
  querying the database can detect the privacy violation.
  \end{abstract}

\maketitle

\section{Introduction}\label{sec:intro}

{I}{n} its most basic form, the scenario we consider can
be described as follows. On one side we have a provider, Bob, who
controls an ordered classical database composed of $N=2^n$ memory
cells. Each cell of the data-base contains an $m$ bit string, so that
the database consists of $N$ strings $A_0, A_1, \cdots, A_{N-1}$. On
the other side, we have the person querying the database, Alice, who
wants to recover the string associated with a memory cell (say the
$j$-th one) but at the same time does not want Bob to know which cell
she is interested in ({\em user privacy}). In a purely classical
setting the simplest strategy for Alice consists in placing a large
number of decoy queries, i.e. she ``hides'' her query among a large
number $M-1$ of randomly selected queries. In this case, she will be
able to get the information she is looking for, while limiting Bob's
intrusion in her privacy. [In fact, the mutual information between
Alice's true query $j$ and Bob's estimate of such value is upper
bounded by $\log_2 (N/M) - (M-1)/M \log_2((N-1)/(M-1))$]. The
drawbacks associated with such procedures are evident. First of all,
the method does not allow Alice to check whether Bob is retaining
information on her queries. Moreover, to achieve a high level of
privacy Alice is forced to submit large amounts of fake queries,
increasing the communication cost of the transition: in particular,
absolute privacy is obtained only for $M=N$, i.e. by asking Bob to
send {\em all} his database. This may not be acceptable if the
database is huge or if it is an asset for Bob ({\em data privacy}).

User and data privacy are apparently in conflict: the most
straightforward way to obtain user privacy is for Alice to have Bob
send her the entire database, leading to no data privacy whatsoever.
Conversely, techniques for guaranteeing the server's data privacy
typically leave the user vulnerable~\cite{SPIR}. At the information
theoretical level, this problem has been formalized as the
Symmetrically-Private Information Retrieval (SPIR)~\cite{SPIR}
generalizing the Private Information Retrieval (PIR)
problem~\cite{PIR,ostro} which deals with user privacy alone. SPIR is
closely related to oblivious transfer~\cite{oblivious}, in which Bob
sends to Alice $N$ bits, out of which Alice can access exactly
one--which one, Bob doesn't know. No efficient solutions in terms of
communication complexity~\cite{amba} are known for SPIR. Indeed, even
rephrasing them at a quantum level~\cite{kerend1,kerend2}, the best
known solution for the SPIR problem (with a single database server)
employs $O(N)$ qubits to be exchanged between the server and the user,
and ensures data privacy only in the case of {\em honest users} (i.e.
users who do not want to compromise their chances of getting the
information about the selected item in order to get more). Better
performance is obtained for the case of multiple non-mutually
communicating servers~\cite{PIR} (although the user cannot have any
guarantee that the servers are not secretly cooperating to violate her
privacy), while sub-linear communication complexity is possible under
the some computational complexity assumption, e.g.~\cite{ostro}. PIR
admits protocols that are more efficient in terms of communication
complexity~\cite{PIR,ostro}.

The Quantum Private Queries (QPQ) protocol we have introduced in
Ref.~\cite{NOSTRO} is a cheat sensitive strategy~\cite{hardy} which
addresses both user and data privacy while allowing an exponential
reduction in the communication and computational complexity with
respect to the best (quantum or classical) single-server SPIR protocol
proposed so far. Specifically QPQ provides a method to check whether
or not Bob is cheating and does not need the exchange of the whole
database (i.e. $O(N)$ qubits): in its simplest form it only requires
Bob to transfer two database elements, identified by $O(\log N)$
qubits, for each query. The QPQ protocol is ideally composed by a
preliminary signaling stage where the user and the database provider
exchange some quantum messages (specifically Alice addresses Bob
receiving some feedback from him) and by a subsequent {\em
  retrieval\&check} stage where Alice performs some simple quantum
information processing on the received messages to recover the
information she is interested in and to check Bob's honesty. The QPQ
security relies on the fact that if Bob tries to infer the query Alice
is looking for, she has a nonzero probability of discovering it. Most
importantly, one can verify that the more information Bob gets on
Alice query, the higher is the probability that he will not pass
Alice's honesty test. In this paper we will derive analytical bounds
for such a theoretical trade-off, and we analyze different variants of
the QPQ protocol.

The main idea behind the protocol is the following. Alice submits her
request to Bob using some quantum information carrier, so that she can
either submit a plain query $|j\rangle$ or a quantum superposition of
different queries $\alpha|j\rangle+\beta|j^\prime\rangle$. Alice
randomly alternates superposed queries and non-superposed queries.
Thus, Bob does not know whether the request he is receiving at any
given time is a superposition of queries or not, so that he does not
know which measurement will leave the information carrier unperturbed:
he cannot extract information without risking to introduce a
disturbance that Alice can detect. Bob can, however, respond to
Alice's request without knowing which kind of query was submitted. His
response will be either of the form $|j\rangle|A_j\rangle$ or of the
form
$\alpha|j\rangle|A_j\rangle+\beta|j^\prime\rangle|A_{j^\prime}\rangle$,
where the first ket is the register that Alice had sent him, the
second ket is a register that contains Bob's answer ($A_i$ being the
answer to the $i$th query), and which may be entangled with the first.
From these answers Alice can both obtain the reply to her query and
check that Bob has not tried to breach her privacy.

The main assumption we adopt is that, for each $j$, there exists a
{\em unique} answer string $A_j$ that can be independently checked by
Alice. [This does not prevent different queries from having the same
answer: indeed we do admit the possibility to have $A_j =
A_{j^\prime}$ for $j\neq j^\prime$.] For example, Alice may be asking
Bob the prime factors of one out of $N$ very large integer numbers
(say the RSA collection) which she cannot factorize by herself.
The above requirements can be relaxed (examples will be provided in
Sec.~\ref{sec:soluzione}), but they are useful as they permit a
considerable simplification of the security proof. For the same
reason, we will focus on the simplest version of the QPQ protocol,
where there exists a reference query $0$ (dubbed {\em rhetoric} query)
which has a known standard answer $A_0$. As discussed in
Ref.~\cite{NOSTRO}, this assumption is not fundamental, but it is very
useful since it allows us to minimize the amount of exchanged signals
in the protocol (as a matter of fact alternative versions of the QPQ
protocol with higher security level can be devised which do not employ
the rhetoric query).

The paper is organized as follows. In Sec.~\ref{sec:notations} we
describe the rhetoric version of QPQ in its basic form and introduce
the notation. This is followed by the technical Sec.~\ref{sec:bob}
where we analyze in detail the most general transformations Bob can
perform on Alice's queries. Section~\ref{sec:security} contains the
main result of the paper: here we introduce the trade-off between
Bob's information on Alice's query and the success probability of him
passing her honesty test. In Sec.~\ref{sec:ent} we present some
variations of the QPQ protocol, one of which exploits entanglement as
a resource to strengthen Alice's privacy. Finally in
Sec.~\ref{sec:what} we analyze what happens when relaxing some of the
assumptions adopted in the security proof. In particular we show that
the basic version of the QPQ described here does not guarantee privacy
if the queries have multiple answers, and we point out a possible
solution in Sec.~\ref{sec:soluzione}.

\section{Preliminaries and notation}\label{sec:notations}

\begin{figure}[t]
\begin{centering}\includegraphics[width=1\columnwidth]{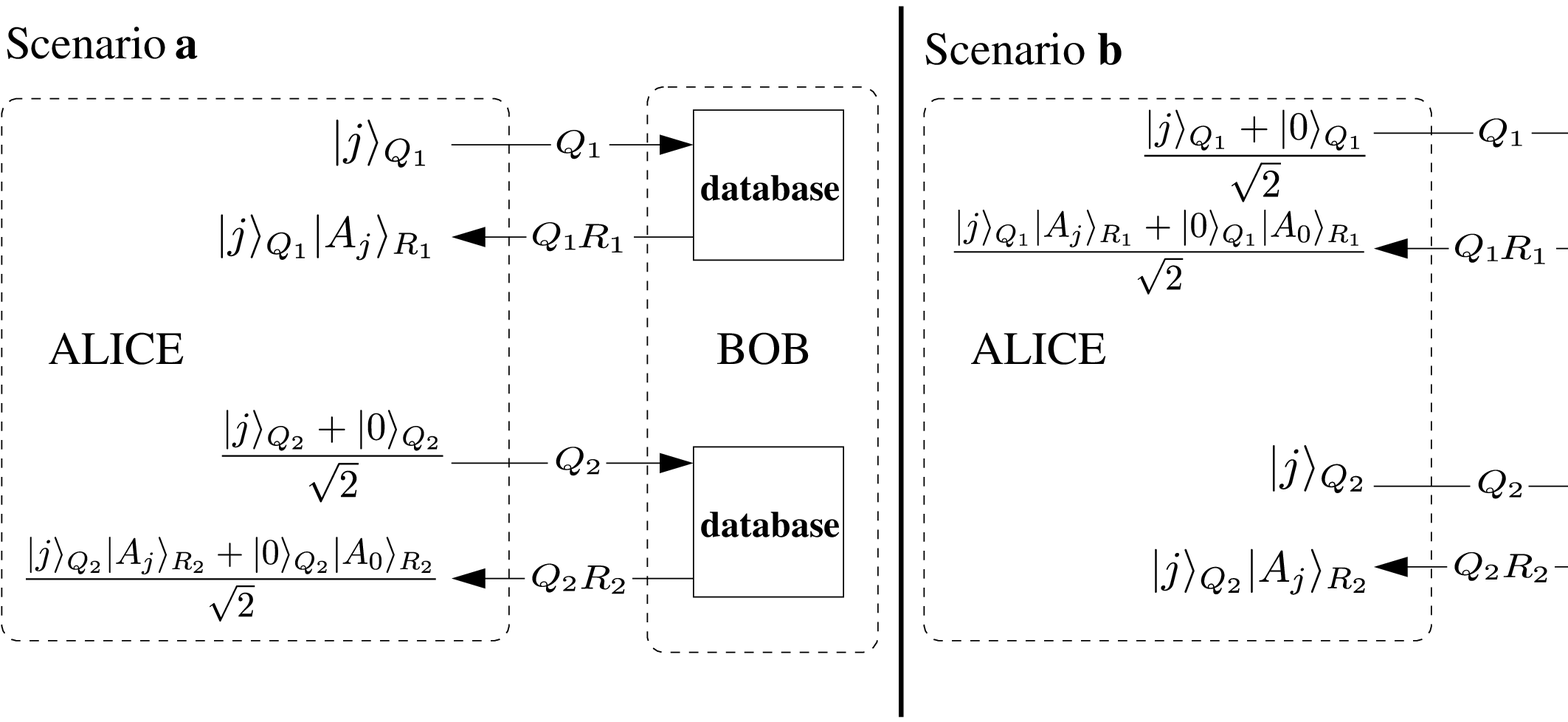}\par\end{centering}
\caption{Scheme of the QPQ protocol with rhetoric questions. Alice
  wants to find out the $j$th record of Bob's database (composed of
  $N=2^n$ records). She then prepares two $n$-qubit registers. The
  first contains the state $|j\rangle_Q$, while the second contains
  the quantum superposition $(|j\rangle_Q+|0\rangle_Q)/\sqrt{2}$
  between her query and the rhetoric question ``$0$'', to which she
  knows the standard answer $A_0$. She then sends, in random order
  (i.e.~randomly choosing either scenario {\bf a} or scenario {\bf
    b}), these two registers to Bob, waiting for his first reply
  before sending the next register. Bob uses each of the two registers
  to interrogate his database using a qRAM device, which records the
  reply to her queries in the two ``reply'' registers $R$. At the end
  of their exchange, Alice possesses the states
  $|j\rangle_Q|A_j\rangle_R$ and
  $(|j\rangle_Q|A_j\rangle_R+|0\rangle_Q|A_0\rangle_R)/\sqrt{2}$,
  where the $A_j$ is the content of the $j$th record in the database.
  By measuring the first she obtains the value of $A_j$, with which
  she can check whether the superposition in the second state was
  preserved. If this is not the case, then she can be confident Bob
  that Bob has violated her privacy, and has tried to obtain
  information on what $j$ was. } \label{f:proto}
\end{figure}
In the rhetoric version of the QPQ protocol (see Fig.~\ref{f:proto})
Alice uses two quantum registers each time she needs to interrogate
Bob's database. The first register contains $|j\rangle$, the address
of the database memory cell she is interested in; the other register
is prepared in a quantum superposition of the type
$(|j\rangle+|0\rangle)/\sqrt{2}$, ``$0$'' being the rhetoric query.
Alice then {\em secretly} and {\em randomly} chooses one of the two
registers and sends it to Bob. He returns the register Alice has sent
to him, together with an extra register in which the corresponding
answer is encoded. In order to reply to Alice's query without knowing
whether it is the superposed query or not, Bob needs to employ the
quantum random access memory (qRAM) algorithm~\cite{NIELSEN,qRAM}.
After Alice has received Bob's first reply, she sends her second
register and waits for Bob's second reply. Again Bob returns her
register together with an extra register which encodes his reply
obtained through a qRAM application. If Bob has followed the protocol
accurately, without trying to extract information, Alice now should
possess a state which encodes the information she is looking for and
an entangled state involving the rhetoric query, whose coherence can
be tested to check Bob honesty, i.e. the two states
$|j\rangle|A_j\rangle$ and
$(|j\rangle|A_j\rangle+|0\rangle|A_0\rangle)/\sqrt{2}$. Alice recovers
the value of $A_j$ by measuring the second register in the first
state, and then she uses this value to prepare a measurement to test
whether the superposition has been retained in the second state
(``honesty test''). Such a measurement is simply a projective
measurement on the state
$(|j\rangle|A_j\rangle+|0\rangle|A_0\rangle)/\sqrt{2}$. If this test
fails, namely if she finds out that the state Bob has sent her back is
orthogonal to the one she is expecting, she can be confident that Bob
has cheated and has violated her privacy. If, instead, the test
passes, she cannot conclude anything. In fact, suppose that Bob has
measured the state and collapsed it to the form $|j\rangle|A_j\rangle$
or to the form $|0\rangle|A_0\rangle$, it still has a probability 1/2
to pass Alice's test of it being of the form
$(|j\rangle|A_j\rangle+|0\rangle|A_0\rangle)/\sqrt{2}$. So Alice's
cheat test allows her to be confident that Bob has cheated if the test
fails, but she can never be completely confident that Bob has not
cheated if the test passes.

We now introduce the notation which will be used. We define $X\equiv
\{ 0,1 , \cdots , N-1\}$ the source space which contains the addresses
$j$ of the memory cells which compose Bob's database, identifying with
$j=0$ the address of the rhetoric query. For each $j$ we define $A_j$
to be the information associated with the $j$-th address. As mentioned
in the introduction the $A_j$ are {\em classical messages} composed of
$m$ bits, and they need not represent distinct messages (i.e. we allow
the possibility that $A_j = A_{j'}$ for $j\neq j'$), but they are
uniquely determined by the value of $j$. In this context, Bob's
database is defined as the ordered set ${\cal D} \equiv \{ A_j | j\in
X\}$ formed by the strings $A_j$. We define $Q= Q_1,Q_2$ the two
quantum registers Alice uses to submit her queries; according to the
protocol, she will first send $Q_1$, wait for Bob's answer and then
send $Q_2$. In this notation, for $k=1,2$, the vector
$|j\rangle_{Q_k}$ is the state of the $k$-th register which carries
the address of the $j$-th database memory. For all $j\neq 0$ we use
the vector $|+j\rangle_{Q_k}$ to represent the superposition of the
$j$-th query and the rhetoric query, i.e.
\begin{eqnarray}
|+j\rangle_{Q_k} \equiv ({|j\rangle_{Q_{k}} 
+ |0\rangle_{Q_{k}}})/{\sqrt{2}}\;, \label{defq1}
\end{eqnarray}
(for $j=0$ we have $|+0\rangle_{Q_k} \equiv |0\rangle_{Q_k}$). We
define $R\equiv R_1, R_2$ the registers on which Bob writes the
information to send back to Alice. After having received $Q_1$ from
Alice, Bob encodes the necessary information on $R_1$ and sends back
to her both $Q_1$ and $R_1$. Analogously, after having received
${Q}_2$, he will encode information on $R_2$ and send her back both
${Q}_2$ and $R_2$. It is useful to also define the vectors
\begin{eqnarray} |C_j\rangle_{Q_1R_1} &\equiv& |j\rangle_{ Q_1}
  |A_j\rangle_{R_1} \;,
\label{check1} \\
|C_{\pm j}\rangle_{Q_kR_k} 
&\equiv& ({|C_j\rangle_{Q_kR_k} \pm |C_0\rangle_{Q_kR_k} })/ {\sqrt{2}}
\label{check} \;,
\end{eqnarray}
(as in Eq.~(\ref{defq1}) for $j=0$ we set $|C_{+ 0}\rangle_{Q_kR_k}
\equiv |C_{ 0}\rangle_{Q_kR_k}$). According to the protocol, the
vectors $|C_j\rangle_{Q_kR_k}$ or $|C_{+j}\rangle_{Q_kR_k}$ are the
states that an honest Bob should send back to Alice when she is
preparing $Q_k$ into the states $|j\rangle_{Q_k}$ or
$|+j\rangle_{Q_k}$, respectively. In fact, the states
$|C_j\rangle_{Q_kR_k}$ and $|C_{+j}\rangle_{Q_kR_k}$ are the result of
the qRAM transformation when it is fed $|j\rangle_{Q_k}$ and
$|+j\rangle_{Q_k}$, respectively. We also introduce an ancillary
system $B$ to represent any auxiliary systems that Bob may employ when
performing his local transformation on the Alice queries, plus
(possibly) an external environment.

Let us use this notation to better formalize the QPQ protocol
described above. Suppose then that Alice wants to address the $j$-th
entry of the database. The protocol goes as follows:

\begin{enumerate}
\item Alice randomly chooses between the two alternative scenarios
  {\bf a} and {\bf b} (see Fig.~\ref{f:proto}). In the scenario {\bf
    a}, she prepares the qubits $Q_1$ in $|j\rangle_{Q_1}$ and the
  qubits $Q_2$ in $|+j\rangle_{Q_2}$. Instead, in the scenario {\bf b}
  she prepares the states $|+j\rangle_{Q_1}$ and $|j\rangle_{Q_2}$.
  This means that, in the scenario {\bf a}, she first sends the plain
  query and then the superposed query. On the contrary, in the
  scenario {\bf b}, she first sends the superposed query and then the
  plain query. Consequently, the input state of the system $QRB$ is
  described by the vectors
\begin{eqnarray}
 |\Psi_j^{(\ell)} \rangle_{QRB} \equiv 
\left\{ 
\begin{array}{l}
|j\rangle_{Q_1}|+j\rangle_{Q_2} 
|000\rangle_{RB}\; \mbox{for $\ell =${\bf a},} \\ \\
|+j\rangle_{Q_1} |j\rangle_{Q_2} 
|000\rangle_{RB}\; \mbox{for $\ell =${\bf b},} \\
\end{array} \right. 
\label{equation1}
\end{eqnarray}
where the index $\ell$ refers to the selected scenario and
$|000\rangle_{RB}$ is the fiducial initial state of the systems
$R=R_1R_2$ and $B$ (it is independent on $\ell$ because Bob does not
know which scenario Alice has chosen).
\item Now Alice sends $Q_1$ and waits until Bob gives her back $Q_1$
  and $R_1$. Then, she sends $Q_2$ and waits until she gets back $Q_2$
  and $R_2$.
\item {\em Honesty Test}: Alice checks the states she has received.
  If she had selected scenario {\bf a}, she performs a von Neumann
  measurement to see if $QR$ is in the state $|C_j\rangle_{Q_1R_1}
  |C_{+j} \rangle_{Q_2R_2}$--- see Eq.~(\ref{check}). Of course, this
  can be done in two steps: first she measures $Q_1R_1$ to learn $A_j$
  and then she uses this value to prepare an appropriate measurement
  on $Q_2R_2$. If the measurement fails, then Alice can definitely
  conclude that Bob was cheating, otherwise she can assume he was
  honest (although she has no guarantee of it). If she had chosen
  scenario {\bf b}, she proceeds analogously, using a von Neumann
  measurement to check if $QR$ is in the state $|C_j\rangle_{Q_2R_2}
  |C_{+j} \rangle_{Q_1R_1}$.
\end{enumerate}

\section{Bob's transformations}\label{sec:bob}
In the QPQ protocol, Alice's privacy relies essentially on the fact
that Bob is not allowed to operate jointly on $Q_1$ and $Q_2$. This a
fundamental constraint: without it, Bob would be able to discover the
index $j$ without Alice knowing it. In fact, the subspaces ${\cal
  H}_j$ spanned by the two vectors $|j\rangle_{Q_1} |+j\rangle_{Q_2}$
and $|+j\rangle_{Q_1} |j\rangle_{Q_2}$ (associated to the two
different scenarios {\bf a} and {\bf b} for the query $j$) are
mutually orthogonal. Thus, such vectors (and then the corresponding
queries) could be easily distinguished by performing on $Q_1Q_2$ a
simple von Neumann measurement defined by the projectors associated
with the spaces ${\cal H}_{j}$. This is a measurement that would allow
Bob to recover Alice's query without disturbing the input states of
$Q_1Q_2$. To prevent this cheating strategy, the QPQ protocol forces
Bob to address $Q_1$ and $Q_2$ separately (i.e.~he has to send the
register $Q_1$ back, before Alice provides him the register $Q_2$).

Bob's action when he receives Alice's first register can be described
by a unitary operator $U_{Q_1RB}^{(1)}$ which acts on the first
register $Q_1$, on $R=R_1R_2$, and on $B$ (and not on the second
register $Q_2$ which is still in Alice's possession). Analogously,
Bob's action when he receives the second register is described by the
unitary operator $U_{Q_2R_2B}^{(2)}$ which acts on $Q_2$, $R_2$, and
$B$ (and not on $Q_1$ and $R_1$ which are now in Alice's possession).
[Note that the above framework describes also the situation in which
Bob is employing non-unitary transformations (i.e. CP-maps), since the
space $B$ can be thought to contain also the Naimark extension that
transforms any CP-map into a unitary.] The above transformations
cannot depend on the selected scenario $\ell$ (as Bob does not know
which one, among $\ell=${\bf a} and $\ell=${\bf b}, has been selected
by Alice). Therefore, within the $\ell$th scenario, the global state
at the end of the protocol is described by the vectors
\begin{eqnarray}\label{finale}
|\Xi^{(\ell)}_j\rangle_{QRB} \equiv U^{(2)}_{Q_2R_2B}
U^{(1)}_{Q_1RB} \; |\Psi^{(\ell)}_j\rangle_{QRB} \;,
\end{eqnarray}
with $|\Psi^{(\ell)}_j\rangle_{QRB}$ given in Eq.~(\ref{equation1}).
\subsection{Some useful decompositions} 
Consider the transformation $U^{(1)}$. In the scenario {\bf a} for
all $j$ we can write
\begin{eqnarray}
  &&\lefteqn{U_{Q_1RB}^{(1)} \; \left( | j\rangle_{Q_1} 
      |000\rangle_{RB} \right)} \label{EECC} \\ \nonumber 
  && = \sqrt{\eta_j^{(1)}} \;| C_j ; \Phi_j^{(1)} \rangle_{Q_1RB} + 
  \sqrt{1- \eta_j^{(1)}} \; |V_j^{(1)}\rangle_{Q_1RB} \;,
\end{eqnarray}
where $| C_j ; \Phi_j^{(1)} \rangle_{Q_1RB}$ stands for the separable
state $| C_j\rangle_{Q_1R_1} |\Phi_j^{(1)} \rangle_{R_2B}$ and where
$|V_j^{(1)}\rangle_{Q_1RB}$ is a vector orthogonal to $|
C_j\rangle_{Q_1R_1}$, i.e.
\begin{eqnarray}
{_{Q_1R_1}\langle} C_j | V_j^{(1)} \rangle_{Q_1RB} = 0\;.
\label{orto}
\end{eqnarray}
With this choice, $\eta_j^{(1)}$ is the probability that the
state~(\ref{EECC}) will be found in $| C_j\rangle_{Q_1R_1}$. In the
scenario {\bf b}, instead, for $j\neq 0$ we can write
\begin{eqnarray}
&&{U_{A_1RB}^{(1)} \big( | + j\rangle_{Q_1} 
|000\rangle_{RB} \big)}\label{EECC1}
 \\ \nonumber 
&&= \sqrt{\overline{\eta}_j^{(1)}} | C_{+j} ;
\overline{\Phi}_j^{(1)} \rangle_{Q_1RB}
+ \sqrt{1- \overline{\eta}_j^{(1)}} |\overline{V}_j^{(1)}\rangle_{Q_1RB}.
\end{eqnarray}
As before $| C_{+j} ; \overline{\Phi}_j^{(1)} \rangle_{Q_1RB} \equiv |
C_{+j} \rangle_{Q_1R_1}| \overline{\Phi}_j^{(1)} \rangle_{Q_1RB}$ and
$|\overline{V}_j^{(1)}\rangle_{Q_1RB}$ is a vector orthogonal to the
``check state'' $| C_{+j}\rangle_{Q_1R_1}$ Alice is expecting, i.e.
\begin{eqnarray}
{_{Q_1R_1}\langle} C_{+j} | \overline{V}_j^{(1)} \rangle_{Q_1RB} = 0\;.
\label{orto1}
\end{eqnarray}
Consequently $\overline{\eta}_j^{(1)}$ is the probability that the
state~(\ref{EECC1}) will pass the test of being in $|
C_{+j}\rangle_{Q_1R_1}$. The state on the first line of
Eq.~(\ref{EECC1}) can be expanded on a basis of which the state on the
first line of Eq.~(\ref{EECC}) is a component. Therefore
$\overline{\eta}_j^{(1)}$ and $\eta_j^{(1)}$ must be related. The
security analysis given in the following sections is based on the study
of this relation. \vskip 1\baselineskip

Analogous decompositions can be given for $U^{(2)}$: in this case,
however, it is useful to describe them not in terms of the input
states, but in terms of the state of the system {\em after it has
  passed the test on the subsystems $Q_1R_1$}. For $j\neq 0$, in the
scenario {\bf a} this gives:
\begin{eqnarray}
 &&{U_{Q_2R_2B}^{(2)} \; \left( | + j\rangle_{Q_2} 
|\Phi_j^{(1)} \rangle_{R_2B} \right) } \label{eecc} \\
&&= \sqrt{\overline{\eta}_j^{(2)}} \; \nonumber
| C_{+j} ; \overline{\Phi}_j^{(2)} \rangle_{Q_2R_2B} + 
\sqrt{1- \overline{\eta}_j^{(2)}} \; |\overline{V}_j^{(2)}\rangle_{Q_2R_2B} \;.
\end{eqnarray}
Here $|\overline{V}_j^{(2)}\rangle_{Q_2R_2B}$ is a vector orthogonal
to $| C_{+j}\rangle_{Q_2R_2}$ of Eq.~(\ref{check}) i.e.
 $${_{Q_2R_2}\langle} C_{+j} | \overline{V}_j^{(2)} \rangle_{Q_2R_2B}
 = 0\;.$$ Thus $\overline{\eta}_j^{(2)}$ is the probability that the
 state~(\ref{eecc}) will pass the test of being in $|
 C_{+j}\rangle_{Q_2R_2}$. Notice that the vector $|\Phi_j^{(1)}
 \rangle_{R_2B}$ in the first line of Eq.~(\ref{eecc}) is the state of
 $R_2B$ one obtains in the scenario {\bf a} if, after the first round,
 the state $Q_1R_1$ passes the test of being $|C_j\rangle_{Q_1R_1}$
 --- see Eq.~(\ref{EECC}). In the scenario {\bf b}, instead, we have
\begin{eqnarray}
 &&{U_{Q_2RB}^{(2)} \left( | j\rangle_{Q_2} \label{eecc1}
|\overline{\Phi}_j^{(1)}\rangle_{R_2B} \right) } \\
\nonumber &&= \sqrt{{\eta}_j^{(2)}} 
| C_{j}; {\Phi}_j^{(2)} \rangle_{Q_2R_2B} + 
\sqrt{1- {\eta}_j^{(2)}} \; |{V}_j^{(2)}\rangle_{Q_2R_2B},
\end{eqnarray}
where $|{V}_j^{(2)}\rangle_{Q_2R_2B}$ is a vector orthogonal to the
state $| C_{j}\rangle_{Q_2R_2}$, i.e.
$${_{Q_2R_2}\langle} C_{j} | {V}_j^{(2)} \rangle_{Q_2R_2B} 
= 0\;,$$ and ${\eta}_j^{(2)}$ is the probability that the
state~(\ref{eecc1}) will be found in $| C_{j}\rangle_{Q_2R_2}$.
\vskip 1\baselineskip

The case $j=0$ has to be treated separately: indeed, if
Alice sends this query then both $Q_1$ and $Q_2$ will be
prepared into $|0\rangle$. In this case, it is then useful to define
$U^{(2)}$ by considering its action on the vector $ | 0\rangle_{Q_2}
|{\Phi}_0^{(1)}\rangle_{R_2B}$ with $|{\Phi}_0^{(1)}\rangle_{R_2B}$
defined as in Eq.~(\ref{EECC}), i.e.
\begin{eqnarray}
&&U_{Q_2RB}^{(2)} \left( | 0 \rangle_{Q_2} 
|{\Phi}_0^{(1)}\rangle_{R_2B} \right) \\
\nonumber &&= \sqrt{{\eta}_0^{(2)}} \;
| C_0 ; {\Phi}_0^{(2)} \rangle_{Q_2 R_2B} + 
\sqrt{1- {\eta}_0^{(2)}} |{V}_0^{(2)}\rangle_{Q_2R_2B},
\label{eecc10}
\end{eqnarray}
where again one has
$${_{Q_2R_2}\langle} C_{0} | {V}_0^{(2)} \rangle_{Q_2R_2B} 
= 0\;.$$

\vskip 1\baselineskip From the above equations, it follows that for
$j\neq 0$ the final state (\ref{finale}), after Bob has finished his
manipulations, can be written as follows for scenario {\bf a}
\begin{eqnarray}
&&|\Xi^{(\mbox{\bf a})}_j\rangle =
\sqrt{\eta_j^{(1)} \overline{\eta}_j^{(2)}}
\; |C_{+j}\rangle_{Q_2R_2} |C_j\rangle_{Q_1R_1} |\overline{\Phi}_j^{(2)}\rangle_B
\nonumber \\
&&+ \sqrt{\eta_j^{(1)}(1-\overline{\eta}_j^{(2)})}
\; |C_j\rangle_{Q_1R_1} |\overline{V}_j^{(2)}\rangle_{Q_2R_2B}
\nonumber 
\\
&&+ \sqrt{1-\eta_j^{(1)}} U^{(2)}_{Q_2R_2B} |+j\rangle_{Q_2} |V_j^{(1)}
\rangle_{Q_1RB} \label{quella},
\end{eqnarray}
where all the terms in the second and third line are orthogonal to
$|C_{+j}\rangle_{Q_2R_2} |C_j\rangle_{Q_1R_1}$.
Analogously, we have for scenario {\bf b}
\begin{eqnarray}
&&|\Xi^{(\mbox{\bf b})}_j\rangle =
\sqrt{\overline{\eta}_j^{(1)} {\eta}_j^{(2)}}
\; |C_{j}\rangle_{Q_2R_2} |C_{+j}\rangle_{Q_1R_1} 
|{\Phi}_j^{(2)}\rangle_B
\nonumber \\
&&+ \sqrt{\overline{\eta}_j^{(1)}(1-{\eta}_j^{(2)})}
\; |C_{+j}\rangle_{Q_1R_1} |{V}_j^{(2)}\rangle_{Q_2R_2B}
\nonumber \\
&&+ \sqrt{1-\overline{\eta}_j^{(1)}} U^{(2)}_{Q_2R_2B} |j\rangle_{Q_2} 
|\overline{V}_j^{(1)} 
\rangle_{Q_1RB} \;, \label{quella1}
\end{eqnarray}
where, again, the states in the last two lines are orthogonal to the
state in the first. Instead, for $j=0$ we have
\begin{eqnarray}
&&|\Xi^{(\mbox{{\bf a},{\bf b}})}_0\rangle =
\sqrt{\eta_0^{(1)} {\eta}_0^{(2)}}
\; |C_{0}\rangle_{Q_2R_2} |C_0\rangle_{Q_1R_1} |{\Phi}_0^{(2)}\rangle_B
\nonumber \\
&&+ \sqrt{\eta_0^{(1)}(1-{\eta}_0^{(2)})}
\; |C_0\rangle_{Q_1R_1} |{V}_0^{(2)}\rangle_{Q_2R_2B} \nonumber \\
&&
+ \sqrt{1-\eta_0^{(1)}} U^{(2)}_{Q_2R_2B} |0\rangle_{Q_2} |V_0^{(1)}
\rangle_{Q_1RB} \label{quellali} \;.
\end{eqnarray}

\section{Information-disturbance tradeoff and
  Privacy}\label{sec:security}

In this section we present an information-disturbance analysis of the
QPQ protocol. This will yield a trade-off which shows that, if Bob
tries to get some information on Alice's queries, then she has a
nonzero probability of detecting that he is cheating. The same
analysis can be easily reproduced for more complicated versions of the
protocol. For instance Alice may hide her queries into superpositions
of randomly selected queries. In this case, the derivation, although
more involved, is a straightforward generalization of the one
presented here.

According to Eq.~(\ref{finale}), to measure Bob's information gain, it
is sufficient to study how the final state of the ancillary subsystem
$B$ depends upon Alice's query $j$. Exploiting the decompositions
introduced in Sec.~\ref{sec:bob} we can then show that one can force
$B$ to keep no track of Alice's query by bounding the success
probabilities that Bob will pass the QPQ honesty test. Specifically,
indicating with $P_j^{(\ell)}$ the success probability associated with
Alice's query $j$ in the $\ell$-th scenario and defining
$\rho_B^{(\ell)}(j)$ the corresponding output state of $B$, in
Sec.~\ref{sec:ancilla} we will prove the following theorem
 
{\bf Theorem:} {\em Choose $\epsilon \in [0,1]$ so that $P_j^{(\ell)}
  > 1 -\epsilon$ for all $j$ and $\ell$. Then
  there exists a state $\sigma_B^*$ of $B$ and a positive constant $c \leq 631$
  such that the fidelities~\cite{fido} $F(\rho_B^{(\ell)}(j) ;
  \sigma_B^*)$ are bounded as follows,
\begin{eqnarray}\label{theo1}
|F(\rho_B^{(\ell)}(j) ; \sigma_B^*) - 1 | < c \; \epsilon^{1/4}\;,
\end{eqnarray}
for all all $j$ and $\ell$. 
}

This implies that, by requiring Bob's probabilities of passing the
honesty test to be higher than a certain threshold $1-\epsilon$, then
the final states of $B$ will be forced in the vicinity of a common
fixed state $\sigma_B^*$, which is independent from the choice of $j$
and $\ell$. This in turn implies that, for sufficiently small values
of $\epsilon$, Bob will not be able to distinguish reliably between
different values of $j$ using the states in his possession at the end
of the protocol. In particular, if $\epsilon = 0$, i.e. if Bob wants
to be sure that he passes the honesty test, then the final states for
any choice of $j$ will coincide with $\sigma_B^*$, i.e.~they will be
completely independent from $j$: he cannot retain any memory of what
Alice's query was. It is also worth noticing that since the total
number of queries, as well as the number of scenarios $\ell$, is
finite and randomly selected by Alice, then the requirement on
$P_j^{(\ell)}$ in the theorem can be replaced by a similar condition
on the {\em average probability} of success\footnote{ Since the number
  $N$ of entries $j$ is finite, the condition $P_j^{(\ell)} > 1
  -\epsilon$ can be imposed by requiring a similar condition on the
  {\em average probability} $P\equiv \sum_{\ell, j} P_j^{(\ell)}
  /(2N)$. Indeed assume $P>1-\epsilon'$ and let $P_*$ be the minimum
  of the $P_j^{({\ell})}$ for all $j$ and $\ell$. Since there are
  $2N$ terms each contributing to $P$ with probability $1/2N$, we have
  $$
  1- \epsilon' < P \leq P_*/2N + (2N-1)/2N$$ that gives $P_* > 1 -2N
  \epsilon'$. The condition $P_j^{(\ell)} > 1 -\epsilon$ then follows
  by taking $\epsilon = 2 N\epsilon'$.}.

In Sec.~\ref{sec:info-dist} we will employ the above theorem to bound
the {\em mutual information} $I$~\cite{CT} that connects the classical
variable $j \in \{ 1, \cdots, N-1\}$, which labels Alice's query, and
Bob's estimation of this variable. Assuming that initially Bob does
not have any prior information on the value of $j$ that Alice is
interested in, we will determine the value $I$ at the end of the
protocol, showing that this quantity is upper-bounded by the parameter
$\epsilon$ of Eq.~(\ref{theo1}). Specifically, we will show that by
requiring that Bob passes the honesty test with a probability greater
than $1-\epsilon$, then Alice can bound Bob's information as 
\begin{eqnarray}
{I}
\leq c \; \epsilon^{1/4} \; \log_2 N \label{ee2212}\;,
\end{eqnarray}
$N$ being the number of database entries: his information is upper
bounded by a quantity that depends monotonically on a lower bound to
his probability $P_j^{(\ell)}$ of passing the honesty test. Thus, if
he wants to pass the honesty test with high probability, he must
retain a low information on Alice's query.

\subsection{Proof of the Theorem}
\label{sec:ancilla}
Assume that Alice randomly chooses the scenarios {\bf a} and {\bf b}
with probability $1/2$. From Eqs.~(\ref{quella}) and (\ref{quella1})
it is easy to verify that the success probability that Bob will pass
the honesty test when Alice is submitting the $j$-th query is
 \begin{eqnarray}
   P_j = \tfrac{1}{2} \left(P_j^{({\mbox{\bf a}})} + P_j^{({\mbox{\bf b}})} \right) = \tfrac{1}{2} \left(
     \eta_j^{(1)} \overline{\eta}_j^{(2)} + \eta_j^{(2)} \overline{\eta}_j^{(1)}\right)\;, \label{proba}
\end{eqnarray}
where $P_j^{({\mbox{\bf a}})} \equiv \eta_j^{(1)}
\overline{\eta}_j^{(2)}$ and $P_j^{({\mbox{\bf b}})} \equiv
\eta_j^{(2)} \overline{\eta}_j^{(1)}$ refer to the success
probabilities in the scenarios {\bf a} and {\bf b}, respectively
(these expressions hold also for $j=0$ by setting
$\overline{\eta}_0^{(1,2)} \equiv \eta_0^{(1,2)}$). The corresponding
output density matrices of the ancillary system $B$ is given by
 \begin{eqnarray}
\rho_B(j) = \frac{1}{2}
\Big[ \rho_B^{\mbox{({\bf a})}}(j) + \rho_B^{\mbox{({\bf b})}}(j) \Big]
\label{eew}
\;,\end{eqnarray}
where, for $\ell=$ {\bf a} and {\bf b}, the state $\rho_B^{(\ell)}(j)$
are obtained by partial tracing on Alice's spaces 
the output vectors of Eqs.~(\ref{quella}) and (\ref{quella1}), i.e.
\begin{eqnarray}
  \rho_B^{(\ell)}(j) &\equiv& \mbox{Tr}_{QR} 
  [ |\Xi_j^{(\ell)} \rangle\langle \Xi_j^{(\ell)}| ] \nonumber \\
  &=& 
  P^{(\ell)}_j \; \sigma_B^{(\ell)}(j) + \left[ 
    1 - P^{(\ell)}_j \right] \;
  \tilde{\sigma}_B^{(\ell)}(j),\label{fff}\\
  \mbox{with}&&\nonumber\\
  \sigma_B^{({\mbox{\bf a}})}(j) &\equiv& 
  |\overline{\Phi}_j^{(2)}\rangle_B\langle \overline{\Phi}_j^{(2)}|\;,
  \\
  \sigma_B^{({\mbox{\bf b}})}(j)& \equiv
  &|{\Phi}_j^{(2)}\rangle_B\langle {\Phi}_j^{(2)}|.
\end{eqnarray}
The quantities $\sigma_B^{(\ell)}(j)$ (for $\ell=$ {\bf a}
and {\bf b}) are the density matrices obtained by projecting
$|\Xi_j^{(\ell)} \rangle_{QRB}$ into the state of $QR$ which allows
Bob to pass the honesty test (i.e. $|C_j\rangle_{Q_1R_1}
|C_{+j}\rangle_{Q_2R_2}$ for $\ell=${\bf a} and
$|C_{+j}\rangle_{Q_1R_1} |C_{j}\rangle_{Q_2R_2}$ for $\ell=${\bf b} ).

In accordance with the theorem's hypothesis, we consider the case in
which the probability of passing the test~(\ref{proba}) for an
arbitrary $j$ is higher than a certain threshold, i.e.
\begin{eqnarray}
P_j^{(\ell)} > 1-\epsilon \label{cond}\;,
\end{eqnarray}
with $\epsilon \in [0,1]$. We will then prove Eq.~(\ref{theo1}) by
identifying the density matrix $\sigma_B^*$ with the pure
$|\Phi_0^{(2)} \rangle$ defined as in Eq.~(\ref{eecc10}) and showing
that the following condition holds
\begin{eqnarray}
F(\rho_B^{(\ell)} (j), |\Phi_0^{(2)}\rangle) > 1- 631 \; \epsilon^{1/4}\;,
\label{re}
\end{eqnarray}
where $F$ is the fidelity~\cite{fido}. Such inequality is a
consequence of the fact that we want Bob to preserve the coherence of
the superposition $|+j\rangle$, and at the same time to answer
correctly to query $|j\rangle$. To derive it we use Eq. (\ref{fff})
and the condition~(\ref{cond}) to write
\begin{eqnarray}
F(\rho_B^{(\ell)}(j), |\Phi_0^{(2)}\rangle)
\label{rei} 
\geq 
(1-\epsilon)\; 
F(\sigma_B^{(\ell)}(j), |\Phi_0^{(2)}\rangle)\;.
\end{eqnarray}
To prove Eq.~(\ref{re}) it is then sufficient to verify that for all $j$ one has
\begin{eqnarray}
  &F(\sigma_B^{(\mbox{\bf a})}(j), |\Phi_0^{(2)}\rangle) = 
  |\langle \Phi_j^{(2)} |\Phi_0^{(2)} \rangle |^2 >1 - 630 \epsilon^{1/4}\;, &\nonumber \\
  &F(\sigma_B^{(\mbox{\bf b})}(j), |\Phi_0^{(2)}\rangle) = |\langle \overline{\Phi}_j^{(2)} |{\Phi}_0^{(2)} \rangle |^2 > 1 - 630 \epsilon^{1/4}.&
\label{reii}
\end{eqnarray}
The derivation is similar to the one used in Ref.~\cite{qsb} and can
be split in two parts, which will be derived in the following:

{\em i)}
First we use Eq.~(\ref{cond}) and
the definitions~(\ref{EECC}) and~(\ref{EECC1})
to show that for 
for all $j\neq 0$ one has
\begin{eqnarray}
&|\langle \Phi_j^{(1)} | \Phi_0^{(1)} \rangle|^2 >
1 - 28 \sqrt{\epsilon}, &
\label{qqqwq}\\ 
&|\langle \overline{\Phi}_j^{(1)} | \Phi_0^{(1)} \rangle|^2 > 
 [ 1 - 2 ( 
2 + \sqrt{2 
\epsilon} ) \sqrt{\epsilon} ]^2 > 1 - 14 \sqrt{\epsilon} .&
\label{qqqwq00}
\end{eqnarray}

{\em ii)} Then we use Eqs.~(\ref{qqqwq}),
(\ref{qqqwq00}) and 
the definitions~(\ref{eecc}) and~(\ref{eecc1})
to verify that for $j\neq 0$ one has
\begin{eqnarray}
&|\langle \Phi_j^{(2)} | \Phi_0^{(2)} \rangle|^2 > 
(1- 315 \;{\epsilon}^{1/4} )^2 > 1 -630 \epsilon^{1/4},&
 \label{qqqwq2} \\ 
&|\langle \overline{\Phi}_j^{(2)} | \Phi_0^{(2)} \rangle|^2 
> (1- 23 \; {\epsilon}^{1/4} )^2 > 1 - 46 \epsilon^{1/4},&
\label{qqqwq1}
\end{eqnarray}
which proves the theorem with $c=630$.

\subsubsection*{Derivation of Part {i)}} 
The condition (\ref{cond}) implies the following inequalities
\begin{eqnarray}
\eta_j^{(1,2)} > 1-\epsilon\;, \qquad \quad \overline{\eta}_j^{(1,2)} 
&>& 1-\epsilon \;, 
\label{hhhh}
\end{eqnarray}
for all $j\in\{0,1,\cdots,N-1\}$. To obtain inequality~(\ref{qqqwq}), we compare Eqs.~(\ref{EECC}) and
(\ref{EECC1}) under the constraint imposed by Eqs.~(\ref{hhhh}). In
particular, we notice that for $j\neq 0$ Eq.~(\ref{EECC}) gives 
\begin{eqnarray}
U_{Q_1RB}^{(1)} \; \left( | + j\rangle_{Q_1} 
|000\rangle_{RB} \right) = |W_j\rangle_{Q_1RB} + |\Delta W_j\rangle_{Q_1RB} ,
\label{EECCJ}
\end{eqnarray}
with 
\begin{eqnarray}
|W_j\rangle
&=&
\tfrac{\sqrt{{\eta}_j^{(1)}} \;| C_{j}; {\Phi}_j^{(1)} \rangle
+ \sqrt{{\eta}_0^{(1)}} \;| C_{0} ;{\Phi}_0^{(1)} \rangle
}{\sqrt{2}} 
\nonumber \\
|\Delta W_j\rangle
&=&
\tfrac{\sqrt{1- {\eta}_j^{(1)}} \; |{V}_j^{(1)}\rangle
+ \sqrt{1- {\eta}_0^{(1)}} |{V}_0^{(1)}\rangle
}{\sqrt{2}}\;, \label{definizioni}
\end{eqnarray}
According to Eq.~(\ref{hhhh}), the vector $|\Delta W_j\rangle_{Q_1RB}$
has a norm of the order
$\epsilon$. This implies that for $\epsilon\ll 1$ the vector (\ref{EECCJ})
almost coincides with $|W_j\rangle_{Q_1RB}$. Analogously,
Eq.~(\ref{EECC1}) and the second inequality of Eq.~(\ref{hhhh}) tell
us that for $\epsilon \ll 1$, the state $ U_{A_1RB}^{(1)}\left( | +
  j\rangle_{Q_1} |000\rangle_{RB} \right)$ almost coincides with the
vector $| C_{+j} ; \overline{\Phi}_j^{(1)} \rangle_{Q_1RB}$.
Combining these two observations, it follows that for $\epsilon\ll1$
the vectors $|W_j\rangle_{Q_1RB}$ and $| C_{+j} ;
\overline{\Phi}_j^{(1)} \rangle_{Q_1RB}$ almost coincide. According
to definition~(\ref{check}), this implies that
$|\Phi_j^{(1)}\rangle_{R_2B}$, $|\Phi_0^{(1)}\rangle_{R_2B}$ and
$|\overline{\Phi}_j^{(1)}\rangle_{R_2B}$ must converge for
$\epsilon\rightarrow 0$. To make this statement quantitatively
precise, evaluate the scalar product between Eqs.~(\ref{EECCJ}) and
(\ref{EECC1}), and obtain the identity
\begin{eqnarray}
  1 &=& \sqrt{\overline{\eta}_j^{(1)}} \label{fffq}
  \langle W_j | C_{+j}; \overline{\Phi}_j^{(1)}\rangle + 
  \sqrt{1- \overline{\eta}_j^{(1)}} 
  \langle W_j |\overline{V}_j^{(1)}\rangle \\
  && + \sqrt{\overline{\eta}_j^{(1)}} 
  \langle \Delta W_j | C_{+j}; \overline{\Phi}_j^{(1)}\rangle + 
  \sqrt{1- \overline{\eta}_j^{(1)}} 
  \langle \Delta W_j |\overline{V}_j^{(1)}\rangle.
  \nonumber \end{eqnarray}
It can be simplified by using the following inequalities:
\begin{eqnarray}
&|\langle W_j | C_{+j} ; \overline{\Phi}_j^{(1)}\rangle|
\leq
 \tfrac{\; \sqrt{\eta_j^{(1)}} 
\; \big|\langle \Phi_j^{(1)} | \overline{\Phi}_j^{(1)} 
\rangle\big| +
\sqrt{\eta_0^{(1)}}\; \big|\langle \Phi_0^{(1)} 
| \overline{\Phi}_j^{(1)} 
\rangle\big|}{2},& \nonumber \\
&|\langle W_j |\overline{V}_j^{(1)}\rangle| 
\leq 1, &\nonumber \\
&|\langle \Delta W_j | C_{+j}; \overline{\Phi}_j^{(1)}\rangle| 
< \sqrt{\epsilon}, \qquad |\langle \Delta W_j |\overline{V}_j^{(1)}\rangle| 
< \sqrt{2 \epsilon}, &\nonumber
\end{eqnarray}
which can be easily derived from Eq.~(\ref{definizioni}) by invoking
the orthogonality conditions~(\ref{orto1}). Replacing the above
expressions into Eq.~(\ref{fffq}) we get
\begin{eqnarray}
1 &<& 
 \tfrac{\; \sqrt{\overline{\eta}_j^{(1)} \eta_j^{(1)}}}{2} 
\; \big|\langle \Phi_j^{(1)} | \overline{\Phi}_j^{(1)} 
\rangle\big| + \tfrac{\; \sqrt{\overline{\eta}_j^{(1)} 
\eta_0^{(1)}}}{2}\; \big|\langle \Phi_0^{(1)} 
| \overline{\Phi}_j^{(1)} 
\rangle\big| \nonumber \\
&&+ 
2 \;\sqrt{\epsilon} +
\sqrt{2}\; \epsilon
\label{fffqo}\;,
\end{eqnarray}
which implies 
\begin{eqnarray}
\big| \langle \Phi_0^{(1)} | \overline{\Phi}_j^{(1)} 
\rangle\big| &>& 
1 - 2 \left( 2 + \sqrt{2 \;
\epsilon} \right) \sqrt{\epsilon} \;,
\label{xx} \\
\big| \langle \Phi_j^{(1)} | \overline{\Phi}_j^{(1)} 
\rangle\big| &>& 1 - 
2 \left( 
2 + \sqrt{2 \;
\epsilon} \right) \sqrt{\epsilon} 
\;.
\label{xx1}
\end{eqnarray}
We are almost there: indeed Eq.~(\ref{xx}) coincides with
Eq.~(\ref{qqqwq00}). To derive Eq.~(\ref{qqqwq}) we apply the
triangular inequality to the vectors $|\Phi_0^{(1)}\rangle$,
$|\Phi_j^{(1)}\rangle$ and $|\overline{\Phi}_j^{(1)}\rangle$.

\subsubsection*{Derivation of Part ii)} The main difference between
the set of Eqs.~(\ref{EECC}), (\ref{EECC1}) and the set of
Eqs.~(\ref{eecc}), (\ref{eecc1}) is the fact that, in the former,
$U^{(1)}$ acts on vectors with fixed $RB$ component, while, in the
latter, $U^{(2)}$ operates on vectors whose $R_2B$ components may vary
with $j$. We can take care of this by replacing
$|\Phi_j^{(1)}\rangle_{R_2B}$ and
$|\overline{\Phi}_j^{(1)}\rangle_{R_2B}$ with the constant vector
$|\Phi_0^{(1)}\rangle_{R_2B}$. This is, of course, not surprising,
given the inequalities of Eqs.~(\ref{qqqwq}) and~(\ref{qqqwq00}). To
see it explicitly, evaluate the scalar product between
$U_{Q_2RB}^{(2)} \left( | j\rangle_{Q_2}
  |\overline{\Phi}_j^{(1)}\rangle_{R_2B} \right)$ and $
U_{Q_2RB}^{(2)} \left( | j\rangle_{Q_2} |{\Phi}_0^{(1)}\rangle_{R_2B}
\right)$. For $j\neq 0$ it gives,
\begin{eqnarray}
\langle \overline{\Phi}_j^{(1)} 
|{\Phi}_0^{(1)}\rangle = 
 \sqrt{{\eta}_j^{(2)}} \; \langle C_{j}; {\Phi}_j^{(2)} | 
U^{(2)} | j ;{\Phi}_0^{(1)}\rangle \nonumber \\
 + \sqrt{1- {\eta}_j^{(2)}} 
\langle {V}_j^{(2)} |
U^{(2)} | j ; {\Phi}_0^{(1)} \rangle \;.
\label{lanb}
\end{eqnarray}
From the inequalities (\ref{hhhh}) and (\ref{qqqwq00}), it then
follows that the modulus of $ \kappa_j \equiv \langle C_{j}; {\Phi}_j^{(2)} | U^{(2)} | j; 
{\Phi}_0^{(1)}\rangle$ must be close to one, i.e. 
\begin{eqnarray}
|\kappa_j| 
 > 1 - (5+ 2 \sqrt{2\epsilon}) \sqrt{\epsilon} 
>1- 8 \sqrt{\epsilon}\;. \label{giggo}
\end{eqnarray}
Proceeding analogously for the vectors $U^{(2)} | +j ;
{\Phi}_0^{(1)}\rangle$ and $U^{(2)} | C_{+j};
\overline{\Phi}_j^{(2)}\rangle$, we obtain
\begin{eqnarray}
|\overline{\kappa}_j| \equiv |\langle C_{+j}\; \overline{\Phi}_j^{(2)} | 
U^{(2)} | + j \;{\Phi}_0^{(1)}\rangle| 
>1- 29 \sqrt{\epsilon}\;. \label{ggggg}
\end{eqnarray}
For all $j\neq 0$ we can then write the following decompositions:
\begin{eqnarray}
 U_{Q_2RB}^{(2)} ( | j\rangle_{Q_2} \label{neecc1}
|{\Phi}_0^{(1)}\rangle_{R_2B}) &=& \kappa_j 
| C_{j} ;{\Phi}_j^{(2)} \rangle_{Q_2 R_2 B} \\ 
&&+ \nonumber
\sqrt{1- |\kappa_j|^2} |{Z}_j\rangle_{Q_2R_2B} \\
 U_{Q_2R_2B}^{(2)} ( | + j\rangle_{Q_2} 
|\Phi_0^{(1)} \rangle_{R_2B} ) &=& \overline{\kappa}_j
| C_{+j} \label{neecc}
; \overline{\Phi}_j^{(2)} \rangle_{Q_2 R_2 B} \\ \nonumber &&+ 
\sqrt{1- |\overline{\kappa}_j|^2} \; |\overline{Z}_j\rangle_{Q_2R2B} \;,
\end{eqnarray}
where $|Z_j\rangle$ and $|\overline{Z}_j\rangle$ are vectors
orthogonal to $| C_{j} ;{\Phi}_j^{(2)} \rangle$
and $| C_{+j} ;\overline{\Phi}_j^{(2)} \rangle$
respectively.
The inequality~(\ref{qqqwq1}) can now be derived by taking the scalar
product between Eqs.~(\ref{neecc}) and (\ref{eecc10}), remembering
that $ |V_0^{(2)}\rangle$ is orthogonal to $|C_0\rangle$ and using
Eqs.~(\ref{hhhh}) and~(\ref{ggggg}). To derive Eq.~(\ref{qqqwq2}),
instead, we first evaluate the scalar product between
Eqs.~(\ref{neecc1}) and (\ref{neecc}) obtaining
\begin{eqnarray}
| \langle \overline{\Phi}_j^{(2)} 
|{\Phi}_j^{(2)}\rangle| > 1 - 
\sqrt{ 2} [ 4 + 5 \sqrt{58} ] \epsilon^{1/4} > 1 - 60 \; \epsilon^{1/4} \;,
\nonumber 
\end{eqnarray}
and then we impose the triangular inequality between the vectors $|
\overline{\Phi}_j^{(2)}\rangle$, $|{\Phi}_j^{(2)}\rangle$ and
$|{\Phi}_0^{(2)}\rangle$.

\subsection{A bound on Bob's information} \label{sec:info-dist}

Here we give an upper bound to Bob's information on the variable $j$.
This can be done by noticing that we can treat $B$ as a {\em quantum
  source} which encodes the classical information produced by the
classical random source $X$. Specifically, this quantum source will
be characterized by the quantum ensemble ${\cal E}\equiv \{ p_j=1/N,
\rho_B(j)\}$, where $p_j=1/N$ is Alice's probability of selecting the
$j$-th query, and $\rho_B(j)$ is given by Eq.~(\ref{eew}). We can
then give an upper bound to Bob's information by considering the
mutual information $I$ associated with the ensemble ${\cal E}$. From
the Holevo bound~\cite{holevo}, we obtain
\begin{eqnarray}
I \leq \chi({\cal E}) \equiv S(\rho_B) - \tfrac{1}{N}\sum_{j=0}^{N-1} {S(\rho_B(j))}
\label{ee}\;,
\end{eqnarray}
where $\rho_B \equiv \sum_{j=0}^{N-1} \rho_B(j)/N$ is the average
state of $B$, assuming that each of Alice's queries is equiprobable.
To simplify this expression, it is useful to express $\rho_B(j)$ as
\begin{eqnarray}
\rho_B(j) = P_j \; \sigma_j + (1- P_j) \;\tilde{\sigma}_j
\label{qq}\;,
\end{eqnarray}
where 
$P_j $ is 
the average probability~(\ref{proba}) 
that Bob will pass the test while Alice is sending
the $j$-th query, and where $\sigma_j$ and $\tilde{\sigma}_j$
are the density matrices
\begin{eqnarray}
&&\sigma_j \equiv( P_j^{(\mbox{\bf a})} \; \sigma_j^{(\mbox{\bf a})} 
 + P_j^{(\mbox{\bf b})}\; \sigma_j^{(\mbox{\bf b})})/{ (2 P_j)} \label{simga} \;,\\
&&\tilde{\sigma}_j \equiv [ (1-P_j^{(\mbox{\bf a})}) \; \tilde{\sigma}_
j^{(\mbox{\bf a})} 
 + (1-P_j^{(\mbox{\bf b})})\; \tilde{\sigma}_j^{(\mbox{\bf b})}]/[
2 (1-P_j)] \;.
 \nonumber 
\end{eqnarray}
This allows us to
write also
\begin{eqnarray}
\rho_B&=& P \; 
\sigma + (1- P) \; \tilde{\sigma}
\label{simga1} \;,\\
\mbox{with}&&\nonumber\\
\sigma &\equiv& \sum_{j=0}^{N-1} \tfrac{P_j}{NP} \sigma_j\;,\
\tilde{\sigma} \equiv \sum_{j=0}^{N-1} \tfrac{1-P_j}{N(1-P)} \tilde{\sigma}_j,
\end{eqnarray}
where $P \equiv \sum_j P_j/N$ is Bob's average probability of passing
the honesty test, which, according to Eq.~(\ref{cond}), must be greater
than $1-\epsilon$. Equations~(\ref{qq}) and (\ref{simga1}) can then
be exploited to produce the following inequalities~\cite{NIELSEN}
\begin{eqnarray}
S(\rho_B) &\leq& H_2(P) + P \; S(\sigma) + (1-P) \; S(\tilde{\sigma}) \;,
\nonumber \\
S(\rho_B(j)) &\geq& P_j \; S(\sigma_j) + (1-P_j) \; S(\tilde{\sigma}_j) \;,
\end{eqnarray}
where $H_2(x) \equiv -x \log x - (1-x) \log (1-x)$ is the binary
entropy. Therefore Eq.~(\ref{ee}) gives
 \begin{eqnarray}
I \leq H_2(P) + P \; \chi( \{ \tfrac{P_j}{NP};
\sigma_j \}) 
+ (1-P) \; \chi( \{ \tfrac{1-P_j}{N(1-P)};
\tilde{\sigma}_j \}) ,\nonumber
 \end{eqnarray}
where $\chi( \{ \frac{1-P_j}{N(1-P)}; \tilde{\sigma}_j \})$ is
the Holevo information associated with a source characterized by
probabilities $\tfrac{1-P_j}{N(1-P)}$. This quantity can never be
bigger than $\log_2 N$ (the same applies to $\chi( \{ \tfrac{P_j}{NP};
\sigma_j \})$, but we are not going to use it). Therefore, we can write
 \begin{eqnarray}
I 
\leq H_2(P) + P \; \chi( \{ \tfrac{P_j}{NP};
\sigma_j \}) + (1-P) \log_2 N,
\label{ee1bis}
\end{eqnarray}
which shows that, in the limit in which $P\rightarrow 1$, the upper
bound is only given by $\chi( \{ \frac{P_j}{NP}; \sigma_j \})$. The
claim is that for $P\sim 1$ this quantity vanishes. Indeed, according
to Eq.~(\ref{reii}) we know that for $\epsilon \rightarrow 0$ the
density matrices $\sigma_j$ converge to the fixed state $|\Phi_0^{(2)}
\rangle_B$, hence
\begin{eqnarray}
\lim_{P\rightarrow 1} \chi( \{ \frac{P_j}{NP};
\sigma_j \}) = \chi( \{ \frac{1}{N};
|\Phi_0^{(2)} \rangle\}) = 0 \;.
\end{eqnarray}
More generally, we now show that $I$ can be bounded from above to any
value $>0$ for $P$ sufficiently close to $1$.

In order to exploit the above relations to give a bound on $I$, let us
introduce the probabilities 
\begin{eqnarray}
q_j \equiv \langle \Phi_0^{(2)} | \sigma_j | \Phi_0^{(2)} \rangle 
\geq 1- 630\; \epsilon^{1/4}\;, \\
q \equiv \langle \Phi_0^{(2)} | \sigma | \Phi_0^{(2)} \rangle
= \sum_{j=0}^{N-1} \tfrac{P_j}{NP} \; q_j 
\geq 1- 630\; \epsilon^{1/4}\;, \label{ecco1}
\end{eqnarray}
(the inequalities simply follow from Eq.~(\ref{reii})). We can then
write
\begin{eqnarray}
\sigma_j &=& q_j\; | \Phi_0^{(2)} \rangle \langle \Phi_0^{(2)}| 
+ (1-q_j) \; \tau_j + \Delta_j \;,\nonumber \\
\sigma &=& q\; | \Phi_0^{(2)} \rangle \langle \Phi_0^{(2)}| 
+ (1-q) \; \tau + \Delta\;,
\end{eqnarray}
where $\tau_j$ are density matrices formed by vectors
$|v_\perp\rangle$ orthogonal to $| \Phi_0^{(2)} \rangle$, $\Delta_j$
are traceless operators containing off-diagonal terms of the form $|
\Phi_0^{(2)} \rangle \langle v_\perp|$, and $\tau \equiv\sum_j P_j
\tau_j/(NP)$. We now introduce a {\em unital} completely positive
trace preserving (CPT) map ${\cal T}$ which destroys the off-diagonal
terms $| \Phi_0^{(2)} \rangle \langle v_\perp|$ while preserving the
corresponding diagonal terms,
and observe that the von Neumann entropy always increases
under the action of a unital map~\cite{STRAETER}. Therefore,
\begin{eqnarray}
 \chi( \{ \tfrac{P_j}{NP};
\sigma_j \}) &\leq& S(\sigma) \leq S({\cal T}(\sigma)) 
\nonumber \\
&=& \nonumber 
S( q\; | \Phi_0^{(2)} \rangle \langle \Phi_0^{(2)}| 
+ (1-q) \; \tau) \\ &\leq& 
H_2(q) + (1-q) S(\tau) \;.\label{eef}
\end{eqnarray}
Now, since $\tau$ is a density matrix in $B$, the quantity $S(\tau)$
can always be upper bounded by $\log_2 d_B$ with $d_B$ the dimension
of $B$. This is not very useful, as $d_B$ can be arbitrarily large.
However, a better solution can be obtained. Indeed, we can show that
the following inequality holds:
\begin{eqnarray}
S(\tau) \leq \log_2 (2N) \;. \label{ep}
\end{eqnarray}
To verify this, we note that the ensemble $\{ \frac{P_j}{NP}; \sigma_j
\}$ is composed by $N$ density matrices of the form~(\ref{simga})
where $\sigma_j^{({\mbox{\bf a}})}$ and $\sigma_j^{({\mbox{\bf b}})}$
are pure vectors satisfying the conditions given in Eq.~(\ref{reii}).
For small $\epsilon$, these $2 N$ vectors are parallel: therefore,
there exists a partial isometry ${\cal I}$ connecting $B$ to a Hilbert
space $B'$ of dimension $2 N$ which maintains their relative distances
intact. Applying such an isometry to all elements of $\{
\frac{P_j}{NP}; \sigma_j \}$ we obtain a new ensemble $\{
\frac{P_j}{NP}; \sigma_j' \}$ of $B'$, whose elements satisfy to the
same relations as the original one. In particular, the two ensembles
possess the same value of $\chi$ (in fact, $\chi$ is an entropic
quantity, whose value depends only on the relations among the ensemble
elements), i.e.
\begin{eqnarray}
\chi(\{ \tfrac{P_j}{NP};
\sigma_j \}) = \chi(\{ \tfrac{P_j}{NP};
\sigma_j' \}) \;,\end{eqnarray}
We can now apply to $\chi(\{ \frac{P_j}{NP};
\sigma_j' \})$ the inequalities (\ref{eef}): the only difference being that now
$\tau$ is a density matrix of $B'$ and hence it satisfies the condition~(\ref{ep}).
Therefore, we can conclude
\begin{eqnarray}
 \chi( \{ \tfrac{P_j}{NP};
\sigma_j \}) &\leq& 
H_2(q) + (1-q) \log_2 (2 N) 
\;. \nonumber
\end{eqnarray}
Replacing this into (\ref{ee1bis}), we finally find
 \begin{eqnarray}
I \leq H_2(P) + P \; H_2(q) + (1-q) + (2 -P-q) \log_2 N 
\label{eewwwewee}\;.
\end{eqnarray}
which thanks to Eq.~(\ref{ecco1}), for sufficiently large $N$ yields
Eq.~(\ref{ee2212}). This means that Alice can limit Bob's information
$I$, by employing in her tests a value of $\epsilon$ sufficiently
small.

\section{QPQ variants and entanglement assisted QPQ} \label{sec:ent}
In this section we discuss few variants of the QPQ protocol that can
be used to improve the security. In particular we introduce an
entanglement assisted QPQ in which Alice entangles her registers
$Q_{1,2}$ with a local ancilla before sending them to Bob. As before,
we will focus for simplicity on rhetoric versions of such variants,
even though similar considerations can be applied also to other
(non-rhetoric) QPQ versions.

An example of cheating strategy will allow us to put in evidence the
aspects of QPQ that these variants are able to improve. Specifically,
suppose that Bob performs a projective measure on all of Alice's
queries to determine the value of the index $j$. As we have seen in
the previous section, he will be by necessity disturbing Alice's state
in average, so that she will have some finite probability to find out
he is cheating. However, if she had chosen scenario {\bf a} [see
Eq.~(\ref{equation1})], then Bob's first measurement on ${Q}_1$ will
return $j$. Now, suppose that his second measurement on Alice's
second request ${Q}_2$ returns the value ``0'' (this happens with
probability $1/2$), then Bob will know that Alice had chosen scenario
{\bf a} and that her query was $j$. In this particular case, he will
be able to evade detection if he re-prepares the system ${Q}_2$ in the
state $|+j\rangle_{Q_2}$. [Of course, this does not mean that he will
evade detection in general, as this is a situation that is
particularly lucky for him, but that has only a small chance of
presenting itself. ] A simple variant of the QPQ protocol can be used
to reduce the success probability of this particular cheating strategy
and in general to strengthen the security of the whole procedure. It
consists in allowing Alice to replace the superposition
$|+j\rangle_{Q_k}$ with states of the form $(|j\rangle_{Q_k} + e^{i
  \theta} |0\rangle_{Q_k} )/\sqrt{2}$, the phase $\theta$ being a
parameter randomly selected by Alice. Since Bob does not know the value
of $\theta$, it will be clearly impossible for him to re-prepare the
correct reply state after his measurement: as a result his probability
of cheating using the simple strategy presented above will be
decreased\footnote{Analogous improvements are obtained by allowing
  Alice to replace the states $|+j\rangle$ with superposition of the
  form $\alpha |j\rangle + \beta |0\rangle$ with the complex
  amplitudes $\alpha$ and $\beta$ being randomly selected.}.
Furthermore since for each given choice of $\theta$, the results of
Sec.~\ref{sec:security} apply, one expects that the use of randomly
selected $\theta$s will result in a general security enhancement of
the QPQ protocol.

In the previous example, the parameter $\theta$ is a {\em secret
  parameter} whose value, unknown to Bob, prevents him from sending
the correct answers to Alice. Another QPQ variant employs entanglement
to enhance security. Suppose that, instead of presenting Bob with the
states $|j\rangle_{{Q}_k}$ and
$|+j\rangle_{{Q}_k}=(|j\rangle_{{Q}_k}+|0\rangle_{{Q}_k})/\sqrt{2}$,
as requested by the QPQ protocol, Alice uses the states
\begin{eqnarray}
  |j\rangle_{{Q}_k}\mbox{ and }
  |\wedge j\rangle_{{Q}_kA}\equiv
  \frac 1{\sqrt{2}}\left[
    |j\rangle_{{Q}_k}|0\rangle_A+
    |0\rangle_{{Q}_k}|j\rangle_A
  \right]
\label{entq}\;,
\end{eqnarray}
where the system $A$ is an ancillary system that Alice does not hand
over to Bob. The protocol now follows the same procedure as the
``canonical'' QPQ described previously, but employing the state
$|\wedge j\rangle$ in place of the state $|+j\rangle$. Of course,
Alice's honesty test must be appropriately modified, as she has to
test whether Bob's actions have destroyed the entanglement between the
ancillary system $A$ and the ${Q}_k$ register.
The main difference
with the canonical QPQ is that here half of the times Bob has only
access to a part of an entangled state: he is even more limited in
re-preparing the states for Alice than in the canonical QPQ. It is
easy to see that the security proof given in the previous sections can
be straightforwardly extended to this version of the protocol, and
that the security bounds derived above still apply: indeed they can be made
even more stringent as Bob has only a limited capacity in his
transformations on Alice's queries, since he does not have access to
the ancillary space $A$. In the situations in which the information carriers employed in the
queries can be put in a superposition of traveling in different
directions~\cite{fabio}, this version of the protocol can easily be
reduced to the canonical QPQ by simply supposing that Alice is in the
possession of the database element $j=0$ corresponding to the rhetoric
question, while, obviously, Bob is in possession of all the remaining
database elements.

\section{What if Alice cannot check the answer to her queries
  independently from Bob?}
\label{sec:what} In deriving the QPQ protocol it is assumed that for
each query $j\in \{1,2,\cdots,N\}$ there is a {\em unique} possible
answer $A_j$ (notice however that two distinct queries can have the
same answer --- i.e. $A_j$ can coincides with $A_{j^\prime}$). One
way to enforce this condition in a realistic scenario is to admit the
possibility that Alice can independently verify the answer that Bob is
sending to her\footnote{This scenario is similar to the one connected
  to NP problems: the function $j \rightarrow A_j$ is difficult to
  compute for Alice, but once a solution is given to her, she can
  easily check if it is correct.}. In this section we will show that
if this is not the case then the basic structure of QPQ does not
prevent Bob to cheat without being discovered by Alice. In
Sec.~\ref{sec:soluzione} we will discuss how one can overcome these
limitations, at least temporarily, by allowing Alice (or third parties
that collaborate with her) to reiterate her query at random times.

\subsection{Successful cheating strategies for a database with
  multiple valid answers}
Here we drop the above hypothesis and give two examples of successful
cheating strategies that allow Bob to spy on Alice's query, and still
pass the honesty test with probability $1$.

\subsubsection{Successful cheating for the rhetoric version of QPQ,
  for databases with multiple valid answers}
Let us start by considering the case of a database with $N=3$ possible
entries in which both the query $j=1$ and the query $j=2$ admit two
distinguishable answers. In particular let $A_1^{(+)}$, $A_1^{(-)}$ be
the answers for $j=1$ and $A_2^{(+)}$, $A_2^{(-)}$ those for $j=2$.

Now, suppose that the unitary $U^{(1)}_{Q_1RB}$ of Eq.~(\ref{finale})
that Bob applies to $Q_1RB$ performs the following mapping
\begin{eqnarray}
|0\rangle_{Q_1} | 0\rangle_{R_1} | 0 \rangle_{B} &\rightarrow& 
|0\rangle_{Q_1} | A_0\rangle_{R_1} | 0 \rangle_{B}\;, \nonumber \\ 
|1\rangle_{Q_1} | 0\rangle_{R_1} | 0 \rangle_{B} &\rightarrow& 
|1\rangle_{Q_1}\tfrac{ |A_1^{(+)} \rangle_{R_1} | + 1 \rangle_{B} + 
|A_1^{(-)} \rangle_{R_1}|- 1 \rangle_{B} }{\sqrt{2}} \;,\nonumber \\
|2\rangle_{Q_1} | 0\rangle_{R_1} | 0 \rangle_{B} &\rightarrow& 
|2\rangle_{Q_1}\tfrac{ |A_2^{(+)} \rangle_{R_1} | + 2 \rangle_{B} + 
|A_2^{(-)} \rangle_{R_1}| - 2 \rangle_{B}}{\sqrt{2}} \;,\nonumber
   \label{ddds}
\end{eqnarray}
where $|A_0\rangle$ is the answer to the rhetoric query and where
\begin{eqnarray}
| \pm 1 \rangle_B \equiv \frac{|0\rangle_B \pm |1\rangle_B}{\sqrt{2}} \;,
\qquad 
| \pm 2 \rangle_B \equiv \frac{|0\rangle_B \pm |2\rangle_B }{\sqrt{2}
}\;,
\label{defff}
\end{eqnarray}
with $|0\rangle_B, |1\rangle_B$ and $|2\rangle_B$ being orthonormal
states of Bob's space $B$. Analogously define $U^{(2)}_{Q_2R_2B}$ as
the unitary operator which performs the following transformation $
|0\rangle_{Q_1}|0\rangle_{R_2} |\psi \rangle_{B} \rightarrow
|0\rangle_{Q_2} |A_0\rangle_{R_2} |\psi \rangle_B$ for all
$|\psi\rangle_B$ of $B$ and
\begin{eqnarray}
 |1\rangle_{Q_1}|0\rangle_{R_2} |\pm 1 \rangle_{B} 
&\rightarrow& |1\rangle_{Q_2} |A_1^{(\pm)}\rangle_{R_2} |\pm 1\rangle_B\;, \nonumber \\
 |2\rangle_{Q_1}|0\rangle_{R_2} |\pm 2 \rangle_{B} 
&\rightarrow& |2\rangle_{Q_2} |A_2^{(\pm)}\rangle_{R_2} |\pm 3
\rangle_B \;. \label{llll}
\end{eqnarray}
According to the above assumptions, if Alice's query is the rhetoric
one (i.e $j=0$) the final state~(\ref{finale}) of the QPQ protocol is
$ |0 \rangle_{Q_2} | A_0\rangle_{R_2} |0\rangle_{Q_1}
|A_0\rangle_{R_1} | 0\rangle_B$. In this case Bob passes the test and
gets $|0\rangle_B$ as output state. For $j=2,3$, instead, we have two
possibilities. In the scenario $\ell=${\bf a} the final state will be
\begin{eqnarray}
&\tfrac{
|j\rangle_{Q_2} | A_j^{(+)} \rangle_{R_2} + |0\rangle_{Q_2}
|A_0\rangle_{R_2} }{2} \;
|j\rangle_{Q_1} | A_j^{(+)} \rangle_{R_1} 
|+ j \rangle_B& \nonumber \\ \nonumber 
&+ \tfrac{|j\rangle_{Q_2} | A_j^{(-)} \rangle_{R_2} + |0\rangle_{Q_2}
|A_0\rangle_{R_2}}{2}
|j\rangle_{Q_1} | A_j^{(-)} \rangle_{R_1} \; |- j \rangle_B \;,&
\end{eqnarray}
while in the scenario $\ell=${\bf b} it will be
\begin{eqnarray}
&\tfrac{
|j\rangle_{Q_1} | A_j^{(+)} \rangle_{R_1} + |0\rangle_{Q_1}
|A_0\rangle_{R_1} }{{2}} 
|j\rangle_{Q_2} | A_j^{(+)} \rangle_{R_2} \; 
| + j \rangle_B &\nonumber \\ \nonumber 
&+ \tfrac{
|j\rangle_{Q_1} | A_j^{(-)} \rangle_{R_1} + |0\rangle_{Q_1}
|A_0\rangle_{R_1} }{{2}} 
|j\rangle_{Q_2} | A_j^{(-)} \rangle_{R_2} \; |- j \rangle_B \;.&
\end{eqnarray}
This means that independently from the selected value of ${\ell}$
Alice will receive the answer $A_j^{(+)}$ half of the times and the
answer $A_j^{(-)}$ in the other half of the times, while Bob will
always pass the honesty test. Moreover in the case in which Alice
receives the answer $A_j^{(+)}$, Bob will get the state $| + j
\rangle_B$ while in the case in which Alice receives the answer
$A_j^{(-)}$ Bob will get the state $|- j \rangle_B$. In average the
state $B$ is $(|0\rangle_B\langle 0| + |j\rangle_B\langle j|_B)/2$.

In conclusion, using $U^{(1)}$
and $U^{(2)}$ as in the previous paragraphs, 
 Bob will always pass the honesty test. 
 Furthermore the output state
of $B$ he gets at the end of the protocol will be
partially correlated with the query $j$ as follows:
\begin{eqnarray}
\begin{array}{rrrc}
\mbox{Query} &\vline&\qquad &\mbox{output state $B$} \\
\hline
j=0 &\vline & \qquad & |0\rangle_B\langle 0| \\
j=1 &\vline & \qquad & (|0\rangle_B\langle 0|+ 
|1\rangle_B\langle 1|)/2 \\
j=2 &\vline & \qquad & (|0\rangle_B\langle 0|+ 
|2\rangle_B\langle 2|)/2
\end{array}
\label{llla}
\end{eqnarray}
Therefore by performing a simple von Neumann measurement on $B$, Bob
will be able to extract some information on $j$, without Alice having
any chance of detecting it.

Notice that, in the example presented here, Bob's info is limited by
the partial overlap between the states $|0\rangle_B\langle 0|$,
$(|0\rangle_B\langle 0|+ |1\rangle_B\langle 1|)/2$ and
$(|0\rangle_B\langle 0 |+ |2\rangle_B\langle 2|)/2$. However, this is
not a fundamental limitation as one can construct more complex
examples (e.g. databases with more than two possible answers for a
single query) for which the amount of info that Bob acquires on $j$
can be arbitrarily high. It is also important to stress that the
above example can be used also to show that Bob will be able to cheat
also in the case in which Alice adopts QPQ strategies more
sophisticated then the simple rhetoric version discussed in this paper
(e.g. instead of sending superpositions of the form $(|j\rangle +
|0\rangle)/\sqrt{2}$ she sends arbitrary superpositions
$\alpha|j\rangle + \beta |0\rangle$ with $\alpha$ and $\beta$
arbitrary amplitudes that only she knows).

\subsubsection{Successful cheating for the non-rhetoric version of
  QPQ, for databases with multiple valid answers}
Here we analyze how multiple valid answers may affect the performance
of the non-rhetoric version of the QPQ protocol (i.e. where Alice is
not using the rhetoric question $j=0$). We give an example of a
successful cheating strategy for a database with $N=3$ queries. For
the sake of simplicity, we will assume that $j=0,1$ have single
answers $A_0$ and $A_1$ respectively, but that $j=2$ is associated
with two distinguishable answers $A_2^{(\pm)}$. As an example of a
non-rhetoric QPQ protocol we consider the case in which Alice, to get
the information associated with the $j$ query, chooses another query
(say the $j'$-th one) and sends sequentially, in random order, states
of the form $\alpha |j\rangle + \beta|j'\rangle$, $|j\rangle$ and
$|j'\rangle$ ($\alpha$ and $\beta$ being amplitudes that only she
knows).

As in the case of the rhetoric version of the protocol, Bob's action
can be described by unitaries. In this case they are
$U^{(1)}$,$U^{(2)}$ and $U^{(3)}$. Notice that the first acts on
$Q_1RB$ the second on $Q_2R_2R_3B$ and the third on $Q_3R_3B$, with
obvious choice of the notation for the subspaces involved. For our
present purpose, it is sufficient to assume that for $k=1,2,3$,
$U^{(k)}$ acts non-trivially only on $Q_k R_kB$ (this is a particular
instance of the general case). We can also assume that
$U^{(1)}$,$U^{(2)}$ and $U^{(3)}$ are identical. We then define such
operators according to the following rules:
\begin{eqnarray}
U^{(k)}_{Q_kR_kB} ( |j\rangle_{Q_k} |0\rangle_{R_k}|0\rangle_B) &=&
|j\rangle_{Q_k} |A_j\rangle_{Rk}|0\rangle_B \;,\nonumber \\
U^{(k)}_{Q_kR_kB} ( |j\rangle_{Q_k} |0\rangle_{R_k}|2\rangle_B) &=&
|j\rangle_{Q_k} |A_j\rangle_{Rk}|2\rangle_B\;,\nonumber 
\end{eqnarray}
{if $j=0,1$} while, for $j=2$,
\begin{eqnarray}
&&U^{(k)}_{Q_kR_kB} ( |2\rangle_{Q_k} |0\rangle_{R_k}|0\rangle_B) \nonumber \\
&&= \nonumber 
|2\rangle_{Q_k} \big(|A_2^{(+)} \rangle_{R_k} | + 2 \rangle_{B} + 
|A_2^{(-)} \rangle_{R_k}| - 2 \rangle_{B} \big)
/\sqrt{2} \;,\\ \nonumber 
&&U^{(k)}_{Q_kR_kB} ( |2\rangle_{Q_k} |0\rangle_{R_k}|2\rangle_B) =
|2\rangle_{Q_k} |A_2^{(\pm)}\rangle_{Rk}| 2\rangle_B \;,
\end{eqnarray}
where $|\pm 2\rangle_B$ are defined in Eq.~(\ref{defff}). If initial
state of the $B$ is $|0\rangle_B$, one can easily verify that Bob will
always pass Alice's honesty test (no matter which superposition
$\alpha|j\rangle + \beta |j'\rangle$ she is using) and that he can
recover part of the information associated with the query. In this
simple example, for instance, he has a not null probability to
identify the query $j=2$. As before, this counterexample can be easily
generalized and improved.

\section{Possible solutions}\label{sec:soluzione}
The case in which different answers may correspond to the same query
is, of course, quite relevant, so that it is natural to ask if the QPQ
protocol can be modified to apply also to this situation. In this
section we give some methods that allow Alice to foil the cheating
strategies described in the previous section temporarily, for as long
as Bob is expecting further queries. 

In the case in which Alice can independently check how many different
replies correspond to the each query (and which are they), then there
is a simple solution that prevents Bob from cheating: we must require
Bob to provide all possible replies in a pre-established order
(e.g.~alphabetically) when he is presented the $j$th query. In this
way, each query has again a unique {\em composite} answer (composed by
the ordered succession of all the possible answers), so that we are
reduced to the canonical QPQ protocol, and Bob is prevented from
cheating.

If, however, Alice cannot independently establish the number of
different replies to each query, then a different strategy is
necessary. [Note that the security proofs given in Sects.~\ref{sec:bob}
and~\ref{sec:security} cease to apply to this version of the protocol,
although conceivably they may be extended to cover also this
situation.]

First of all, we must require that each of the possible replies to the
$j$th query is uniquely indexed by Bob. This means that there should
be a unique answer to the question ``What is the $k$th possible reply
to the $j$th query?'' Of course, this by itself is insufficient to
guarantee that Bob cannot employ the cheating strategies of the
previous section, as Alice cannot independently check the uniqueness
of Bob's indexing (since she does not know all the possible answers to
the $j$th query). However, she can check whether Bob will always
answer in the same way to repeated queries. From Eqs.~(\ref{llll})
and (\ref{llla}), it follows that, as soon as Bob measures his system
$B$, he might gain information on the value of $j$, but at the same
time he loses information on which (among all the possible answers to
the $j$th query) he had presented to Alice. If he wants to be sure
that he keeps on providing always the same answer to repeated queries
on Alice's part, he must preserve his system $B$ without trying to
extract information from it. He can measure the system $B$ only when
he is confident that Alice will not be asking him the same query
anymore. In a multi-party scenario, we can also think of a situation
where multiple cooperating parties ask Bob the same queries and
compare the replies they receive from him. If they find that his
answers when he is asked the $k$th reply to the $j$th query to do not
match, then they can conclude that he has been cheating: he has not
assigned a unique index to all the possible replies to the $j$th
query, and he has taken advantage of the cheating strategies detailed
in the previous section. 

Bob is thus placed in the awkward situation of possessing information
on Alice's query in the system $B$ entirely in his possession, but of
being prevented from accessing such information. This is a temporary
solution, since, as soon as Bob is certain that he will not be asked
the $j$th query anymore, he can measure the system $B$ and extract the
information stored on it. He is kept honest only as long as he is in
business (and, of course, he is in business only as long as he is
honest).

\section{Conclusions}
In conclusion, we have given a security proof of the QPQ protocol
introduced in~\cite{NOSTRO}. It is based on quantitative
information-disturbance tradeoffs which place an upper bound on the
information Bob can retain on Alice's query in terms of the
disturbance he is producing on the states that he is handing back to
her (see Sects.~\ref{sec:bob} and~\ref{sec:security}). A nonzero
information retained by Bob implies a nonzero disturbance on Alice's
states, which she can detect with a simple measurement (the ``honesty
test''). If the honesty test fails, she can conclude that Bob has
certainly cheated. If, on the other hand, the test passes, she can
tentatively conclude that Bob has not cheated (although she cannot be
certain of it).

In addition, we have given some variants of the protocol to further
increase Alice's security, i.e.~to reduce Bob's probability of evading
detection when cheating. These variants either exploit secret
parameters, or exploit entanglement with an ancillary system Alice
retains in her possession (see Sect.~\ref{sec:ent}).

Finally, we have seen that Bob can successfully cheat without being
detected if we drop the assumption (which is at the basis of the QPQ
protocol) that to each query there can be associated only a single
answer $A_j$ (see Sect.~\ref{sec:what}). In fact, if we assume that
there exist two (or more) different replies $A_j\neq A'_j$ to the
query $j$, then Bob can find out the value of $j$, evading detection
by Alice with certainty. We discussed some strategies that allow Alice
to protect herself also in this situation, at least as long as Bob can
expect further queries from her or from other parties who may
cooperate with her (see Sect.~\ref{sec:soluzione}).

\section*{Acknowledgments}

We acknowledge discussions with E. Kushilevitz, S. Micali, F.
Sciarrino, and M.~Sudan. V.~G.~acknowledges support from the Quantum
Information Research program of Centro di Ricerca Ennio De Giorgi of
SNS. S.~L.~acknowledges fruitful discussions with S.~Brin and
L.~Page.

\end{document}